# "PeriHack": Designing a Serious Game for Cybersecurity Awareness

Roberto Dillon
School of Science & Technology
James Cook University
Singapore
roberto.dillon@jcu.edu.au

Arushi
School of Science & Technology
James Cook University
Singapore
arushi@my.jcu.edu.au

*Abstract*—This paper describes the design process for the cybersecurity serious game "PeriHack". Publicly released under a CC (BY-NC-SA) license, PeriHack is a board and card game for two players or teams that simulates the struggle between a red team (attackers) and a blue team (defenders). The game requires players to explore a sample network looking for vulnerabilities and then chain different attacks to exploit possible weaknesses of different nature, which may include both technical and social engineering exploits. At the same time, it also simulates budget level constraints for the blue team by providing limited resources to evaluate and prioritize different critical vulnerabilities. The game is discussed via the lenses of the AGE and 6-11 Frameworks and was primarily designed as a learning tool for students in the cybersecurity and technology related fields.

Keywords— cybersecurity, serious games, tabletop games, training, awareness, STEM education.

## I. INTRODUCTION

With cybersecurity incidents constantly on the rise in an ever-changing technological landscape [1], the need for related training has never been more important across many different industries. Hence, a variety of approaches to engage different types of crowds, from students to common users to professional practitioners, are required. As stated in [2], the learning process is considered to be most effective in an active, experiential setting when students are required to solve specific problems while receiving appropriate feedback for their actions. Serious Games (SGs) have been used for this purpose for many years and different studies such as [3], [4] and [5] confirmed how being engaged in a game offers participants a much more positive experience than simply being exposed to the same learning materials via a standard, passive lecture.

Educational games have to strike the right balance between learning material and a fun gameplay experience and certain themes may be more suitable for certain settings and purposes, as exemplified in [6]. In the specific domain of Cybersecurity, competition and curiosity, as defined in [7], are two emotional elements that are likely candidate for engaging players in any form of gamified training dedicated to this subject. In fact, these are at the core of the ever-popular red team/blue team confrontations and Capture the Flag (CTF) competitions, respectively, where participants need to overcome each other or discover hidden secrets, which are widespread training scenarios across many conferences and events [8] including well known international venues such as [9].

Gamified educational platforms for cyber defense training have also been designed and adopted at the highest levels by NATO and affiliated governments [10], [11] while more abstract serious games in this area have become increasingly popular in the past few years [12], [13]. Among these, it is noteworthy to mention games that focus on specific issues, such as "Protection Poker" [14] to prioritize risks in software engineering projects or "Elevation of Privilege" [15] to practice threat analysis, but we also have more general, introductory and fun games such Control-Alt-Hack [16], a commercial board game based on the mechanics of an existing game, Ninja Burger [17]. Here multiple players (3 to 6) take the role of a hacker with predefined skills across different missions, competing for becoming the most successful white hat on the board.

PeriHack (short for "Perimeter Hacking") falls into the latter group and aims at recreating a stylized and simplified environment where, nonetheless, players will learn about actual potential issues and weaknesses, articulating an attack across different phases to ultimately gain a better understanding of how cybersecurity attack and defense operations do work in practice.

## II. THE GAME

The proposed board game aims at providing a quick way to reflect and put together different bits of knowledge related to common vulnerabilities and attack vectors by pitting an attacking (red) team against a defending (blue) team. In the game, the former has the objective of breaching the security of a typical company network infrastructure by performing certain actions, such as Distributed-Denial-of-Service (DDOS), SQL injection (SQLi) or even phishing and social engineering attacks, while the latter has a certain budget available to increase defenses by, for example, patching systems, implementing network balancing strategies, training employees and so on. Attacks and defenses are then implemented by corresponding cards that are placed on specific network elements on the board, representing the company's infrastructure. The strength of the resulting attacks and defenses are then compared together with the outcome from rolling a polyhedral 20-sided die, following the same approach common to RPG games for skill related rolls where the die result plus eventual bonuses is compared to a set target that needs to be passed to determine the success or failure of a certain action. In our case, the formula is:

$$D20 + Attack\ Bonuses > 10 + Defense\ Bonuses \qquad (1)$$

In designing PeriHack our aim was twofold. For the red team player(s) we wanted to provide a way to illustrate different phases of an attack, giving the ability to players to connect the dots and understand how a coordinated sequence of actions can lead to actual security breaches. On the other hand, for the blue team, the objective was to make them aware of the countless possible vulnerabilities that can be present in any given infrastructure and realize how protecting a certain area may actually imply the allocation of budget and resources that may have also been needed somewhere else, hence opening attacking opportunities that if, discovered by the attackers, would still make the overall infrastructure at risk.

*A. The Board*

The game is played on a board representing a typical office compound and related network infrastructure (Figure 1).

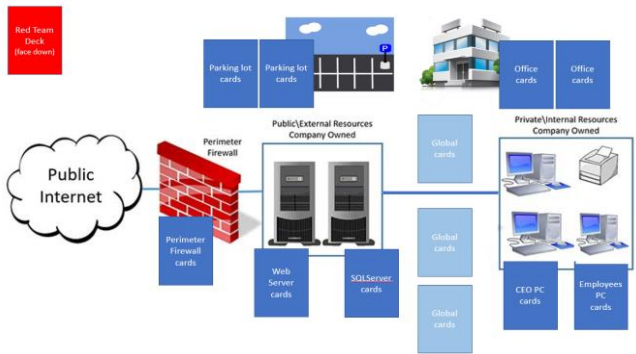

Figure 1: PeriHack board, outlining all the different components representing the targeted company infrastructure. Board, cards and rulebook can be downloaded from github via the following address: https://bit.ly/3kWPolu

The board shows both the physical infrastructure, including premises such as office space and a parking lot, and the main office network, featuring a firewall, a web server, a database, and employees' PCs. Each of these has a dedicated space where corresponding attack and defense cards can be played. The board also allows for three "global defense" cards (GC) to be played. These are cards that potentially have a companywide scope and can enhance the defense of multiple areas, for example, running a staff awareness training program will increase the likelihood of repelling a phishing or a USB drop attack, while enforcing a constant software update policy will reduce the risk of falling for known vulnerabilities related to ransomware, crypto miners, backdoors etc.

*B. The Cards*

Each attack and defense card has a common structure as illustrated in Figure 2. The cards include a type of possible attack that could be exploited (in Attack card) and defense technique (in defense card) with any eventual bonus. Additionally, the card includes its target (i.e. the specific place in the infrastructure where it shall be placed) besides its prerequisite or cost, respectively.

Overall, there are a total of 18 different attack cards available in the attacker's deck (Table 1) where each card has three copies, with the exception of "swap", for which there are five, and the "zero day attack", which is unique, for a total of 54 cards in the deck.

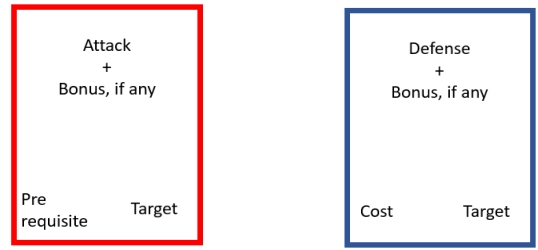

Figure 2: Attack (left) and Defense card template

TABLE I. ATTACKER'S DECK

| Individual Attacker's Cards | | | | | |
|---|---|---|---|---|---|
| Zero Day Attack (*1) | DOS attack via Botnet (*3) | DOS attack via Ping Flood (*3) | Tailgating (*3) | USB Drop (*3) | Man in the Middle (*3) |
| SQLi Attack (*3) | USB Rubber Ducky (*3) | Watering Hole Attack (*3) | Phishing Campaign (*3) | Spear Phishing Campaign (*3) | Stored XSS attack (*3) |
| Reflected XSS attack (*3) | Rogue AP Install (*3) | Ransomware Install (*3) | Backdoor install (*3) | Install Cryptominer (*3) | Swap: Change an existing card for a new random one (*5) |

TABLE II. DEFENDER'S DECKS

| Global Cards Deck (GC) | | | |
|---|---|---|---|
| Ask Advisory Board for additional budget** (*1) | Renew all SW licenses and Patch all SW (*1) | Staff Awareness Training (*1) | Regular offsite backups (*1) |
| **Individual Defense Cards Deck (IC)** | | | |
| Firewall rules to limit ICMP traffic (*1) | Network Load Balancing (*1) | Refactor website to protect against XSS and SQLi (*1) | Add Firewall rules: all 3rd party traffic is untrusted until otherwise verified (*1) |
| Database Honeypot (*1) | Server Honeypot (*1) | Security on Lookout for USB drops (*1) | Security on lookout for Rogue AP (*1) |
| Security on lookout for Tailgating (*1) | Implement 2FA for logins on PCs (*1) | No Additional Defences Here (*5) | |

Defense cards come in two decks: global cards (GC) and individual defense cards (IC). All available defense cards are summarized in Table 2. Cards in the former group can affect multiple board areas and have to be placed face up. Note that there are 4 GC but only 3 slots on the board and, once placed, these cards cannot be stacked/removed/exchanged later in the game, forcing the blue team to define a strategy and take some hard decisions.

Sample cards are presented in figures 3 and 4 for Attack and Defense respectively:

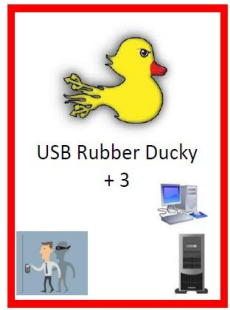

Figure 3**:** The "USB Rubber Ducky" attack card gives a plus 3 attack bonus to the roll and can be played onto PCs or servers, but needs a previously successful tailgating card as a prerequisite to be deployed.

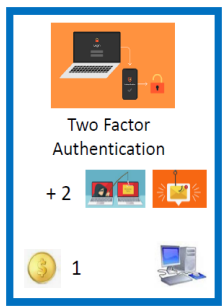

Figure 4: Two Factor Authentication: implementing 2FA costs 1 coin and can be played on the employees' PCs. It adds a +2 bonus defense to phishing and spear phishing attacks.

*C. The Rules*

At the start of the game, the Blue Team begins by allocating its initial budget (e.g. 10 coins) to buy both GC and IC cards. The GCs are placed face-up and the ICs are placed face-down on the respective spots.

At the same time, the Red Team starts with picking five random attack cards from its deck and initially have five coins in its budget that later in the game can allow purchase of additional cards if needed. If any "Swap" card is drawn, this does not count and a new card can be drawn to reach five playable attacking cards.

Based on the five cards, the Red Team can pick one suitable "Winning Condition" card, which include one of the following objectives:

1. DDOS. Can be achieved via:
    a. Successful BOTNET or Ping Flood Attack
2. Database breach. Can be achieved via:
    a. SQLi
    b. Zero Day on DB Server
3. Harvest website users' credentials. Can be achieved via:
    a. Reflected XSS
    b. Stored XSS
4. Harvest employees' credentials. Can be achieved via:
    a. Watering hole attacks
5. Shut down company operations. Can be achieved via:
    a. Install ransomware on server or any PCs
    b. Zero Day on server
6. Spy. Can be achieved via:
    a. Install backdoor on Server or CEO PC
    b. Zero Day on Server or CEO PC
7. Crypto Mining. Can be achieved via:
    a. Gain control of PC (not CEO) to Install malicious SW

As illustrated in the list, each winning condition can be achieved via different attacks, often including multiple phases due to the pre-requisite of each step and corresponding card. The winning objective has also to be kept hidden from the Blue Team and placed face down on the board on the dedicated slot.

Once the Blue team has set their cards and the Red team has drawn their cards and winning condition, the game can begin. In each round the Red team can play one specific attack card plus swap an existing card or buy up to two additional cards by using its available budget. If buying new cards, existing cards also need to be discarded if the overall number of playing cards in the player's hand is equal or greater than five.

Once played, a card has to be discarded but the same attack can be played again in a later round if another card of the same type is available. In this case, though, the defending player will add +1 to its defense roll.

If, by the end of the last round, the winning condition is not achieved, the Blue team is the winner. If, on the other hand, the Red team achieves the condition, it can stop the game, show the card, and explain how the attack was delivered successfully to win the game.

III. GAME ANALYSIS

By analyzing the game via the AGE and 6-11 Frameworks (see [18] and [7]), a descriptive set of tools successfully applied across different serious games projects such as [19],[20] and [21] among others, we can appreciate how the game was designed to engage players via their competitive instincts as well as their desire to find out whether their planning and actions to breach the target would be successful or not (i.e. curiosity), as shown in Figure 5.

At the centre of the gameplay we have the attempts for the red team to breach the target, which is achieved by playing different cards as discussed in Section 2.B.

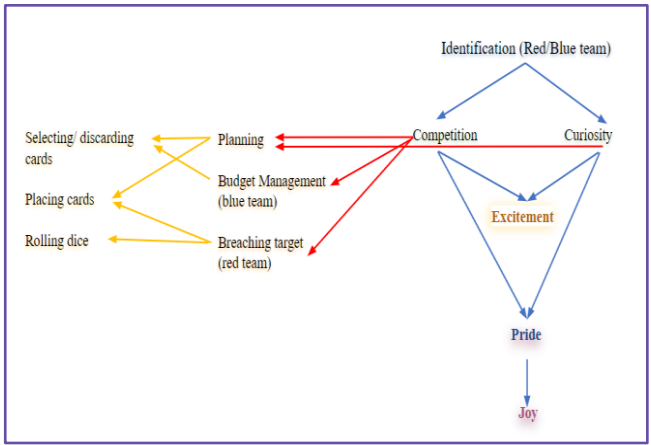

Figure 5: AGE Framework analysis for PeriHack, including all the relevant Actions (left), Gameplay (centre) and Experience (right) elements and their relationships.

To breach the target according to a specific winning condition, it is important to note how the relevant multiple steps may be achieved by different means following alternative paths. Possible ways to chain the attacks are shown in Figure 6.

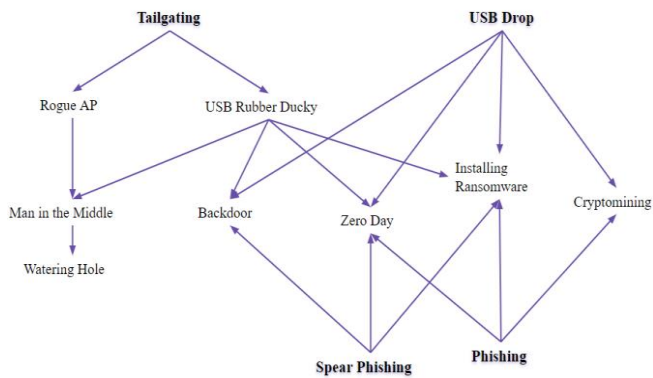

Figure 6: Required steps that need to be chained together for fulfilling a successful attack. Root nodes are in bold. While some attack cards are straightforward and can be played independently (i.e. "stand alone attacks" such as compromising a database via SQLi, DDOS Botnet/Ping flood, Stored/Reflected XSS) others need up to four steps (e.g. harvesting user data via a watering hole attack).

For example, while a winning condition requiring a database breach may be achieved by a single straightforward SQLi attack, a more complex requirement to harvest employees' credentials will require a sequence of well executing steps including for example, a physical infiltration on the company premises to install a rogue AP and establish a man-in-the-middle (MITM) attack. This would then allow the hacker to sniff network traffic and then hijack employees towards a malicious website, effectively implementing a watering hole attack and achieving the original goal.

## IV. Conclusions and Future Work

PeriHack has been designed as an educational game to gain general cybersecurity awareness from both a red and blue team perspectives: the former can obtain a better understanding of how a cyber attack can start and evolve while the latter can appreciate the struggle of managing a multitude of possible weaknesses with limited resources, hence the necessity of taking decisions based on perceived priorities and limited information. Nonetheless, the game can work also as a template for further development: new cards can be added as well as different layouts, to customize the hacking scenario for specific needs and settings.

A proper evaluation of the game is also needed, and is planned for a following study where the game can be tested by following the examples and approaches outlined in [22] and [23].